\documentclass[sigconf]{acmart}
\renewcommand\footnotetextcopyrightpermission[1]{} 
\copyrightyear{2018}
\acmYear{2018}
\setcopyright{acmlicensed}
\acmConference[ACM Hypertext '19]{ACM Hypertext '19}{June 03--05, 2019}{XXX, XXX}
\acmBooktitle{XXX}
\acmPrice{15.00}
\acmDOI{10.1145/1122445.1122456}
\acmISBN{978-1-4503-9999-9/18/06}

\usepackage{times}  
\usepackage{helvet}  
\usepackage{courier}  
\usepackage{url}  
\usepackage{graphicx}  
\usepackage{textcomp}
\usepackage{verbatim}
\usepackage{mathtools}
\usepackage{colortbl}
\usepackage{multirow}
\usepackage{dblfloatfix}
\usepackage{color}
\usepackage{enumitem}
\usepackage[export]{adjustbox}
\usepackage[]{subfig}

\renewcommand{\footnotesize}{\tiny}
\definecolor{LightCyan}{rgb}{0.854, 0.839, 0.803}

\begin{document}

\title{Tracking Temporal Evolution of Graphs using Non-Timestamped Data}

\author{Sujit Rokka Chhetri}
\authornote{All authors contributed equally to this research.}
\affiliation{%
  \institution{University of California Irvine}
}
\email{schhetri@uci.edu}

\author{Palash Goyal}
\authornotemark[1]
\affiliation{%
  \institution{USC Information Sciences Institute}
}
\email{palashgo@usc.edu}

\author{Arquimedes Canedo}
\authornotemark[1]
\affiliation{%
  \institution{Siemens Corporate Technology}
}
\email{arquimedes.canedo@siemens.com}

\begin{abstract}
Datasets to study the temporal evolution of graphs are scarce. To encourage the research of novel dynamic graph learning algorithms we introduce YoutubeGraph-Dyn (available at \url{https://github.com/palash1992/YoutubeGraph-Dyn}), an evolving graph dataset generated from YouTube real-world interactions. YoutubeGraph-Dyn provides intra-day time granularity (with 416 snapshots taken every 6 hours for a period of 104 days),  multi-modal relationships that capture different aspects of the data, multiple attributes including timestamped, non-timestamped, word embeddings, and integers. Our data collection methodology emphasizes the creation of time evolving graphs from non-timestamped data. In this paper, we provide various graph statistics of YoutubeGraph-Dyn and test state-of-the-art graph clustering algorithms to detect community migration, and time series analysis and recurrent neural network algorithms to forecast non-timestamped data. 
\end{abstract}

\maketitle

\section{Introduction}
Graph learning is a growing field tasked with learning representations of networks~\cite{DLGraphs2, DLGraphs1}. Since many real-world problems can be represented with graphs, the are several graph datasets available to the community to test new learning algorithms. However, real-world networks evolve temporally and exhibit dynamic patterns that these datasets fail to capture. Recent work on dynamic graph learning~\cite{DynGEM, dyngraph2vec} has shown that these methods are capable of predicting the evolution of networks. Unfortunately, there are very few real-world dynamic graph datasets available. This includes Hep-th, a collaboration graph of authors in a High Energy Physics Theory conference~\cite{Gehrke}, composed of instances spanning from January 1993 to April 2003; and Autonomous Systems (AS)~\cite{LeskovecAS}, a communication network of who-talks-to-whom from the BGP (Border Gateway Protocol) logs, composed of 733 instances spanning from November 8, 1997 to January 2, 2000. These dynamic graph datasets have some limitations. First, the granularity of the temporal data is greater than a day and therefore it is not possible to study fast dynamics (i.e., by the hour). Second, only one type of relationship is encoded (e.g., who-collaborates-with-whom, who-talks-to-whom) and therefore it is not possible to study related dynamics and their interdependencies. Third, nodes and edges do not provide attributes and therefore these datasets are not suitable for testing deep learning algorithms. 

To accelerate the research on dynamic graph representation learning, this paper introduces \textit{YoutubeGraph-Dyn}, a new dynamic graph dataset generated from the YouTube API that captures different interactions between channels, videos, and comments. YouTube is a global-scale social network where people interact through video views, comments, likes, dislikes, and subscriptions. These multi-modal interactions occur in real time and their rich semantics provides an opportunity to analyze real-world temporal patterns. Youtube Graph-Dyn addresses the limitations of existing dynamic graph datasets as follows:
\begin{itemize}
    \item \textbf{Fine granularity}: graph snapshots every 6 hours.
    \item \textbf{Multi-modal}: encoding various types of relationships between channels and videos through comments, ownership, and communities.
    \item \textbf{Multiple attributes}: including both time-stamped and non-timestamped attributes including word embeddings, lists of word embeddings, and integers.
    \item \textbf{Reproducibility}: to encourage the further advancement of dynamic graph analysis and other deep learning techniques, we make the YoutubeGraph-Dyn dataset available\footnote{YoutubeGraph-Dyn - \url{https://github.com/palash1992/YoutubeGraph-Dyn}} to the community.
\end{itemize}

People in social networks form communities. YoutubeGraph-Dyn provides a ground truth to study the evolution of communities over time. In particular, YoutubeGraph-Dyn's multi-modal interactions motivate new applications in the prediction of user migration to new communities. Similarly, social networks provide platforms to spread ideas. YoutubeGraph-Dyn's multiple attributes motivate new applications in the prediction of content virality, peer influence, and content popularity.

In this paper, we introduce YoutubeGraph-Dyn's key characteristics. To do so, we analyze the predictability of YouTube channels' subscriber, video, and comment counts with different models: autoregressive integrated moving average (ARIMA) models, long short term memory (LSTM), and gated recurrent units (GRU) recurrent neural networks. Communities in YoutubeGraph-Dyn are analyzed through clustering analysis using state-of-the-art graph embedding algorithms~\cite{wang2016structural} on an induced channel-comment-video temporal graph. This graph captures the interactions between communities through comments in YouTube videos.

This paper is organized as follows. Section~\ref{sec:relatedwork} presents the related work to distinguish YoutubeGraph-Dyn's contributions. Section~\ref{sec:datacollection} presents our data collection methodology emphasizing non-timestamped data. Section~\ref{sec:graphdata} presents YoutubeGraph-Dyn's statistics and evaluates community clustering and data forecasting. Section~\ref{sec:discussion} discusses the limitations of YoutubeGraph-Dyn and provides an outlook for future work. Section~\ref{sec:conclusion} concludes.

\section{Related Work}~\label{sec:relatedwork}
YouTube data has been instantiated into various datasets over the years. Most of these datasets have one or more of the limitations discussed in the introduction. 

In~\cite{SFU}, snapshots were taken every 2-3 days. A snapshot is a directed graph, where each video is a node in the graph. If a video $b$ is in the related video list of video $a$, then there is a directed edge from $a$ to $b$. Notice that the related video list is the result of Youtube's recommender system and it is unclear what criteria are used to determine how two videos are related. This dataset only includes commentCount as an attribute, but no actual comments.


The Youtube-D, Youtube-U~\cite{Mislove}, and IMC 2007 datasets~\cite{Mislove, mislove-2007-socialnetworks} were collected to analyze the growth of the user base in the social network as a whole. These datasets consist of user-to-user graphs with daily snapshots between December 10th, 2006 and January 15th, 2007, and February 8th, 2007 to July 23rd, 2007, representing 201 days of growth. This dataset has been also analyzed to identify network communities~\cite{SNAP, LeskovecYoutube}.

Although not suitable for graph learning, but worth mentioning, is the YouTube-8M dataset~\cite{Youtube8M,Youtube8Mpaper}. The YouTube-8M dataset is a multi-label video classification dataset composed of 8 million videos, 500K hours of machine-generated annotations, and a vocabulary of 4800 visual entities. One possible direction for future work would be to induce video-to-annotation, visual entities-to-annotations, or video-to-visual entities temporal graphs.

A multi-attribute dataset very similar to YoutubeGraph-Dyn, but targetted to sentiment analysis and natural language processing, is~\cite{KaggleYoutube}. It consists of daily snapshots of up 200 trending YouTube videos in the United States and the United Kingdom. In comparison, YoutubeGraph-Dyn features more than 6,000 videos.

Dynamic graph representation techniques~\cite{DynGEM, dyngraph2vec, zhu2016scalable, zhang2018timers} have used dynamic graph data available in various domains. 
TIMERS~\cite{zhang2018timers} uses multiple distinct snapshots of Facebook, Wikipedia and DBLP graphs. \textit{dyngraph2vec}~\cite{dyngraph2vec} uses High Energy Physics collaboration networks and Autonomous Systems router networks. The data used in such learning methods are scarcely available and often with a granularity of a few days or months. Furthermore, for many of these networks, the number of snapshots available are not sufficient to learn long temporal dependencies.

This paper aims to introduce a new dynamic data set which captures information about the interactions between channels, videos and comments in YouTube. Our dataset has a granularity of 6 hours enabling its use for more fine-grained temporal analysis.

\section{Data Collection Methodology}\label{sec:datacollection}
To understand the temporal dynamics in the YouTube graph, we need to analyze the Youtube signals that are available through the Youtube API~\cite{YoutubeAPI}. We collected the metadata from 6,342 channels, 277,604 videos, and more than 20 million comments from 2018/07/13 18:00:00 to 2018/10/26 with a frequency of 6 hours (00:00, 06:00, 12:00, and 18:00). These 6,342 channels were selected based on their mentions or their videos mentions on Twitter. The rationale behind this design choice is that content with the potential of becoming popular is being discussed in other social networks beyond YouTube, e.g., on Twitter. We bootstrapped the original channel list from 16,209 YouTube video links posted in Twitter from 2018/07/06 to 2018/07/12. We filtered the Twitter Firehose by the hashtag $\#$youtube and parsed the tweets looking for YouTube URLs.

In this paper, we focus on the three main abstractions in Youtube: (1) a \textit{channel} provides a user a presence in the social network that allows them to upload videos; (2) a \textit{video} is the unit of exchange in Youtube where other users can view, like, dislike, and comment; and (3) a \textit{comment thread} is a collection of comments and comment replies associated to a video. Because the path to fame is a process that occurs in time, we need to distinguish between \textit{timestamped} and \textit{non-timestamped} data.

Timestamped data provides a date; for example, when a comment was made, when a video was uploaded. From timestamped data, we can induce any snapshot of the network by filtering the data by time. Therefore, a single sample suffices to induce any snapshot in time.

Non-timestamped data does not provide an explicit date; for example, a video's current view count, a current channel's number of videos. To associate a date for the non-timestamped data one must sample it explicitly and with a certain frequency to capture changes. To illustrate this problem, consider a channel's subscriber count twenty-four hours ago to be 100 subscribers, eighteen hours ago to be 350, twelve hours ago to be 500, and now to be 200. If sampled daily, we would only see a 2$\times$ growth in subscribers but we would miss the growth from 100 to 500 and the reduction to 200. Thus, capturing fine grain network dynamics requires more frequent sampling compared to what is available in current datasets.

\subsection{Attributes with unique identifiers}
\begin{figure}
 \centering
   \includegraphics[width=1\columnwidth]{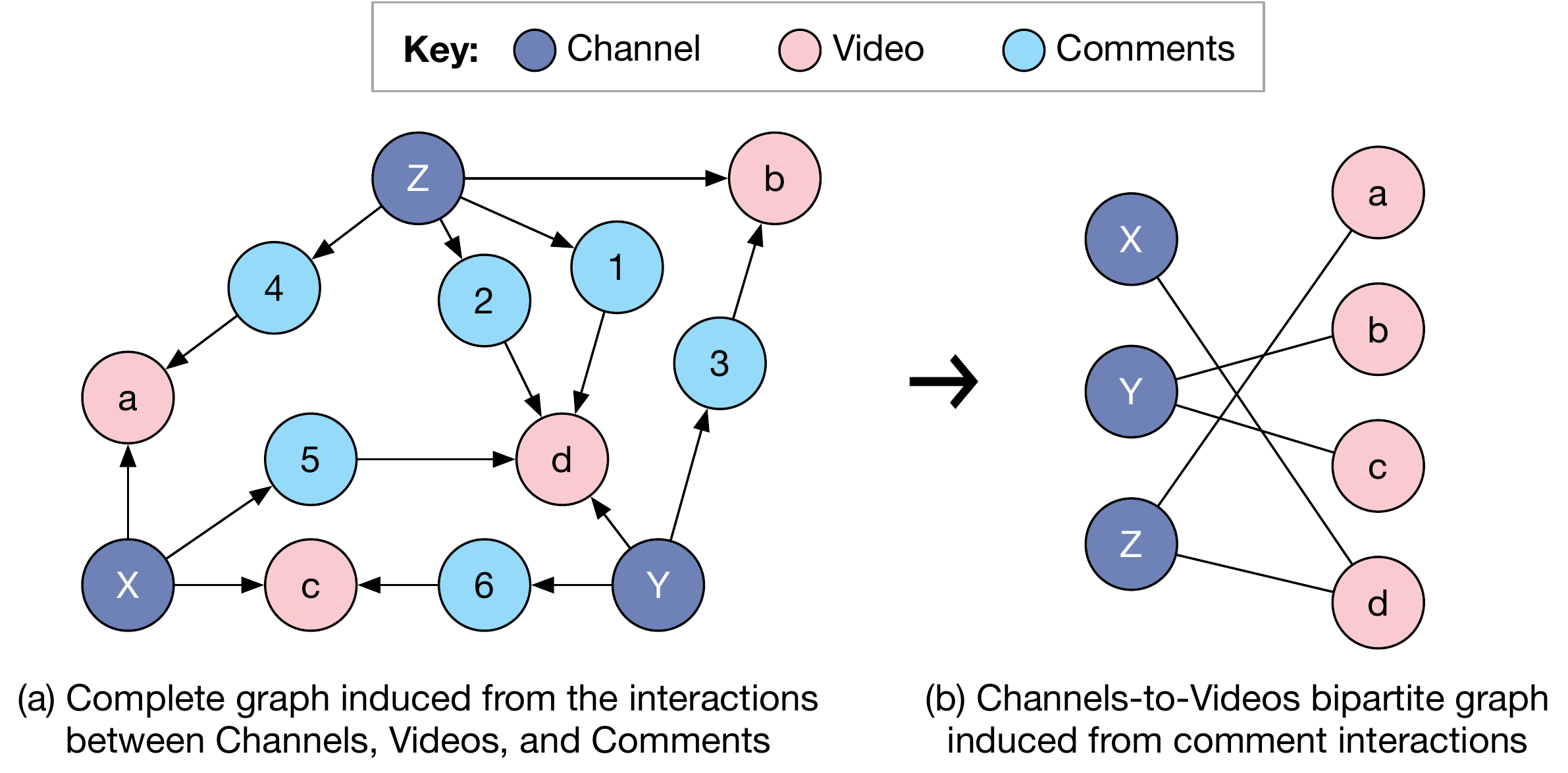}
   \caption{Sample graphs generated from the dataset.}
   \label{fig:socialnetwork}
 \end{figure}

Table~\ref{tab:uniqueid} shows the collected data attributes with a unique identifier (UID) that help us to build the network. Figure~\ref{fig:socialnetwork} shows a sample network induced from the relationships between UIDs. Purple nodes represent channels with their UIDs (X, Y, Z), pink nodes represent videos with their UID (a, b, c, d), and blue nodes represent comments with their UIDs (1, 2, 3, 4, 5, 6). The direction of the arrow represents the originator (source) and the recipient (sink). In Section~\ref{sec:graphdata} we provide the details of the types of graphs that can be induced from this data and analyze the key connectivity metrics in the network.

\begin{table}
\caption{Unique ID attributes.}
\label{tab:uniqueid}
\begin{center}
\begin{tabular}{|l|l|l|}
\hline
\rowcolor{LightCyan}
\textbf{Source} & \textbf{Attribute} & \textbf{Type}\\
\hline
Channel & Channel UID & String\\
 & List of video UIDs & List of strings\\
\hline
Video & Video UID & String\\
 & Parent channel UID & String\\
 & Comment thread UID & List of strings\\
\hline
Comment & Comment thread UID & String\\
Thread & Parent video UID & String\\
 & Parent channel UID & String\\
\hline
\end{tabular}
\end{center}
\end{table}

\subsection{Non-timestamped attributes}
Table~\ref{tab:metadata} shows the collected non-timestamped attributes. For each, we added the snapshot date when they were sampled. Although some attributes such as a channel's title, description, and topic categories (assigned by Youtube) are unlikely to change over time, they may be indicators of popularity. For example, by referencing trending topics, a channel owner may drive more visits to its channel. The same intuition applies to a video's title, description, tags (assigned by the user), duration, and category ID (assigned by Youtube). 

A channel's subscriber count, view count, and video count are expected to change with more frequency and may be stronger indicators of popularity. So are a video's view count, like count, dislike count, and comment count. In Section~\ref{sec:graphdata} we provide a detailed analysis of channels' subscriber count. It is worth mentioning that although we are able to sample a channel's subscriber count, we were not able to collect the subscriber's channel UIDs because this requires explicit authorization from the channel's owner. Access to the subscriber list would provide a more detailed insight into the network. A subscription is a strong indicator of interest among the social network. However, without subscriber lists, the social network can still be analyzed via comments.

Comments are the most frequently changing attribute and are a good indicator of popularity. Channels and videos with more comment activity are likely to be more popular. We capture the comments text as non-timestamped data because users are allowed to modify or delete their comments. Also, a channel owner is allowed to disable comments for a video. These are important events that are captured by our dataset but also present a data collection problem because they cause gaps in the data. 

\begin{table}
\caption{Non-timestamped data attributes.}
\label{tab:metadata}
\begin{center}
\begin{tabular}{|l|l|l|}
\hline
\rowcolor{LightCyan}
\textbf{Source} & \textbf{Attribute} & \textbf{Type}\\
\hline
Channel & Snapshot date & Date\\ 
 & Title & String\\
 & Description & String\\
 & Topic categories & List of strings\\
 & Subscriber count & Int\\
 & View count & Int\\
 & Video count & Int\\
 & Video list & List of videos\\
\hline
Video & Snapshot date & Date\\
 & Title & String\\
 & Description & String\\
 & Tags (user defined) & List of strings\\
 & Duration & Int (seconds)\\
 & Category ID & Int\\
 & View count & Int\\
 & Like count & Int\\
 & Dislike count & Int\\
 & Comment count & Int\\
\hline
Comment & Snapshot date & Date\\
Thread & Comments text & String\\
\hline
\end{tabular}
\end{center}
\end{table}

\begin{figure*}[!ht]
    \centering
    \begin{adjustbox}{minipage=\linewidth,scale=1}
    \subfloat{\includegraphics[width=0.32\textwidth]{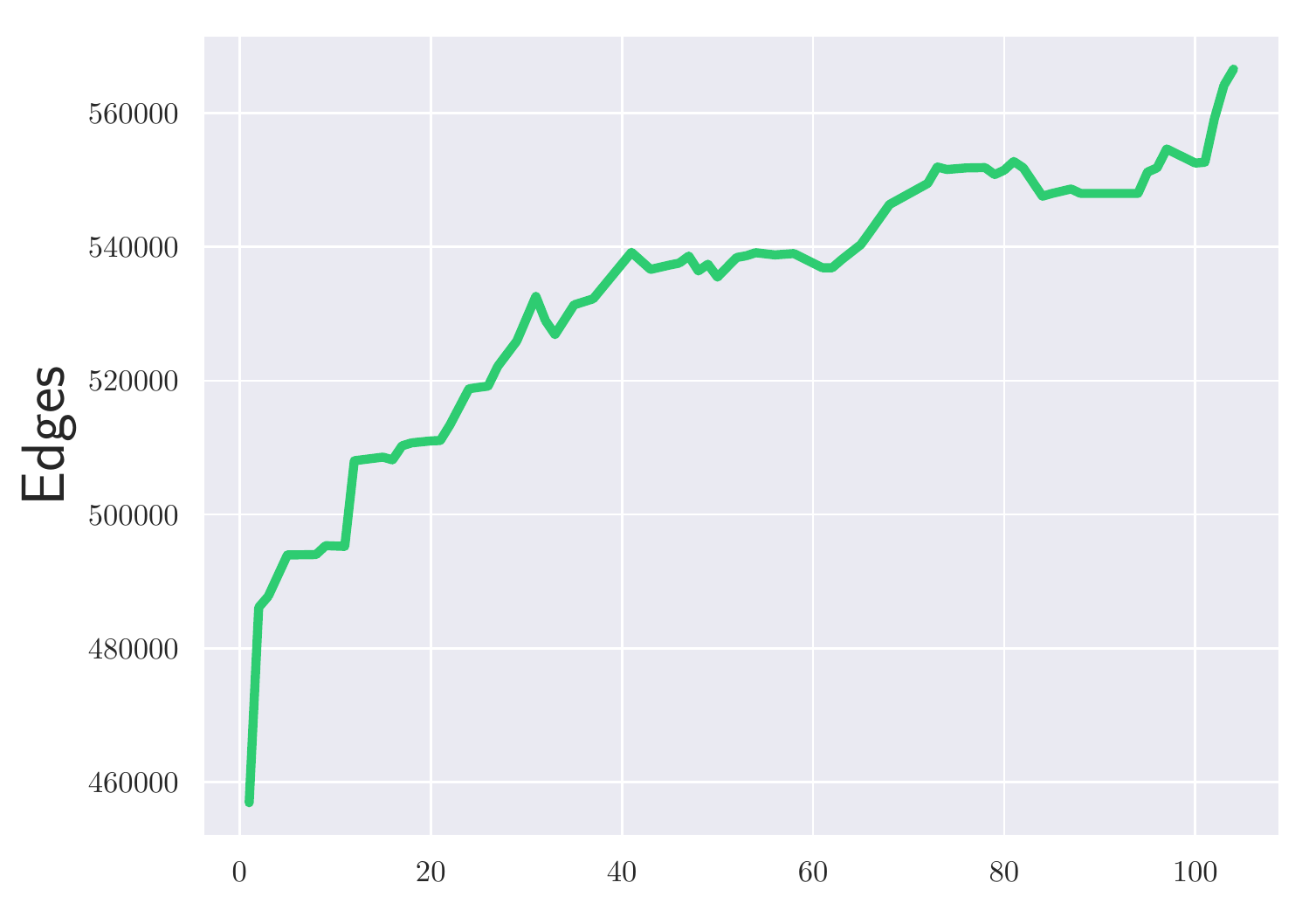}}
    \hfil
    \subfloat{\includegraphics[width=0.32\textwidth]{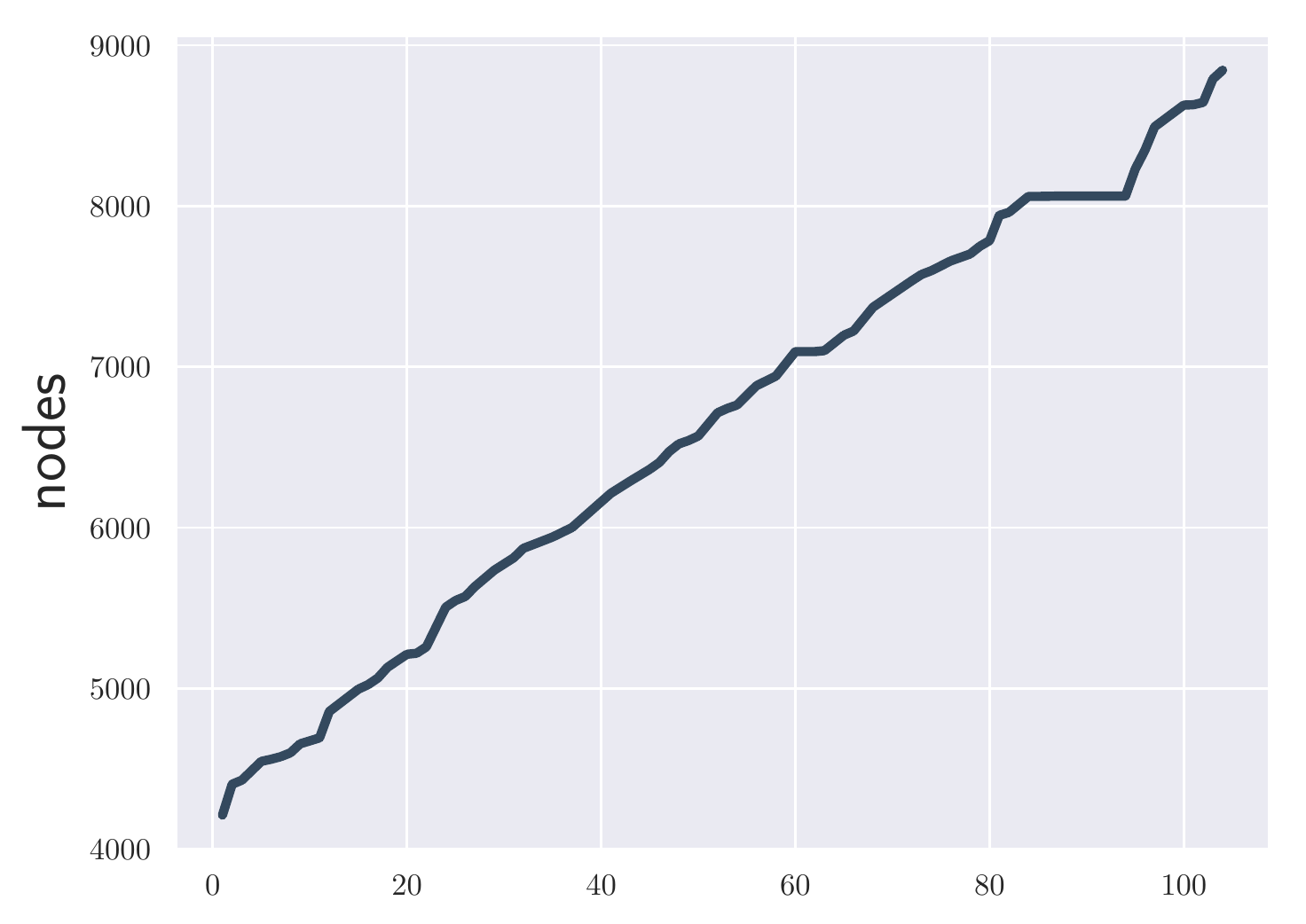}}
    \hfil
    \subfloat{\includegraphics[width=0.32\textwidth]{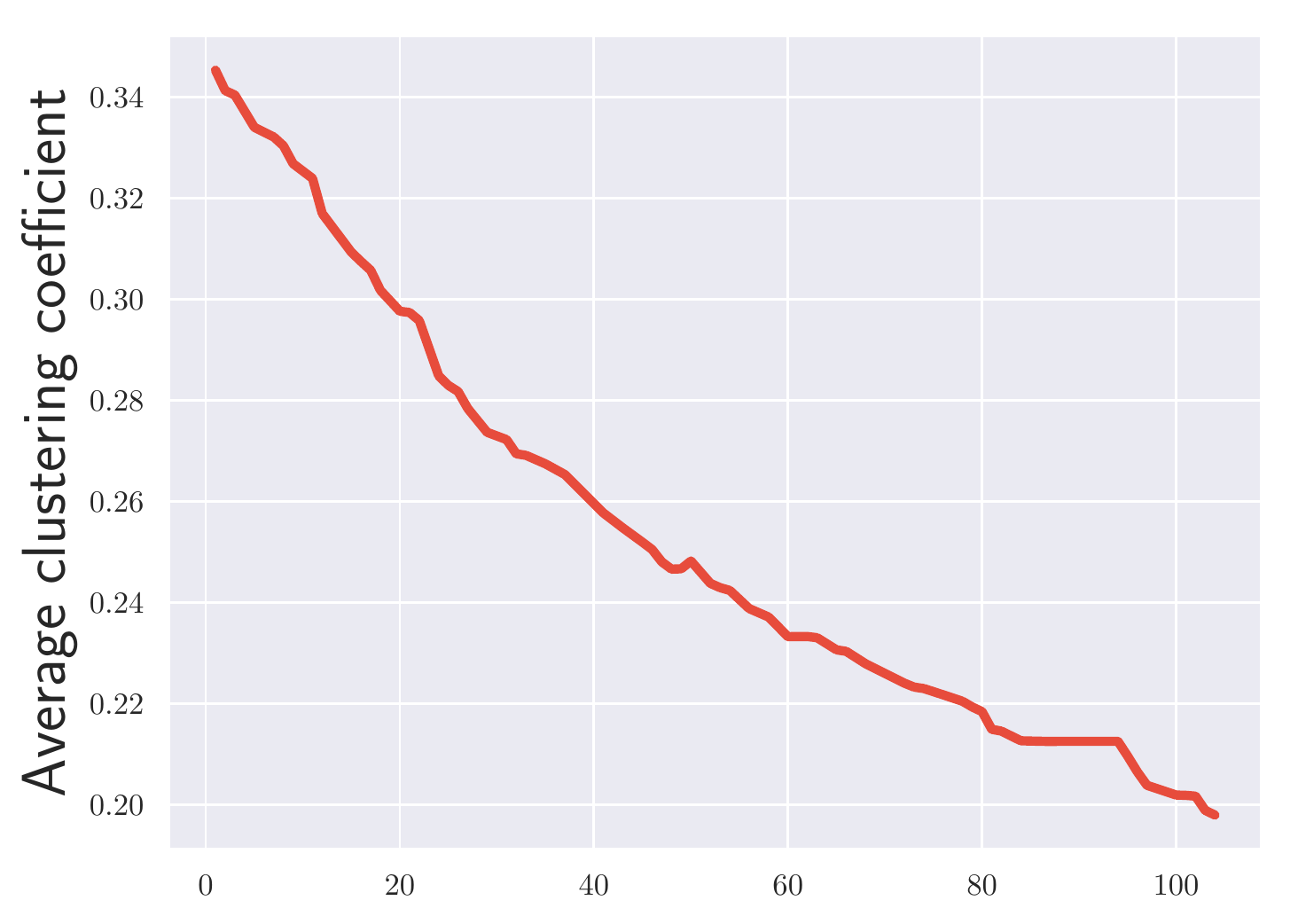}}
      \hfil
     \subfloat{ \includegraphics[width=0.32\textwidth]{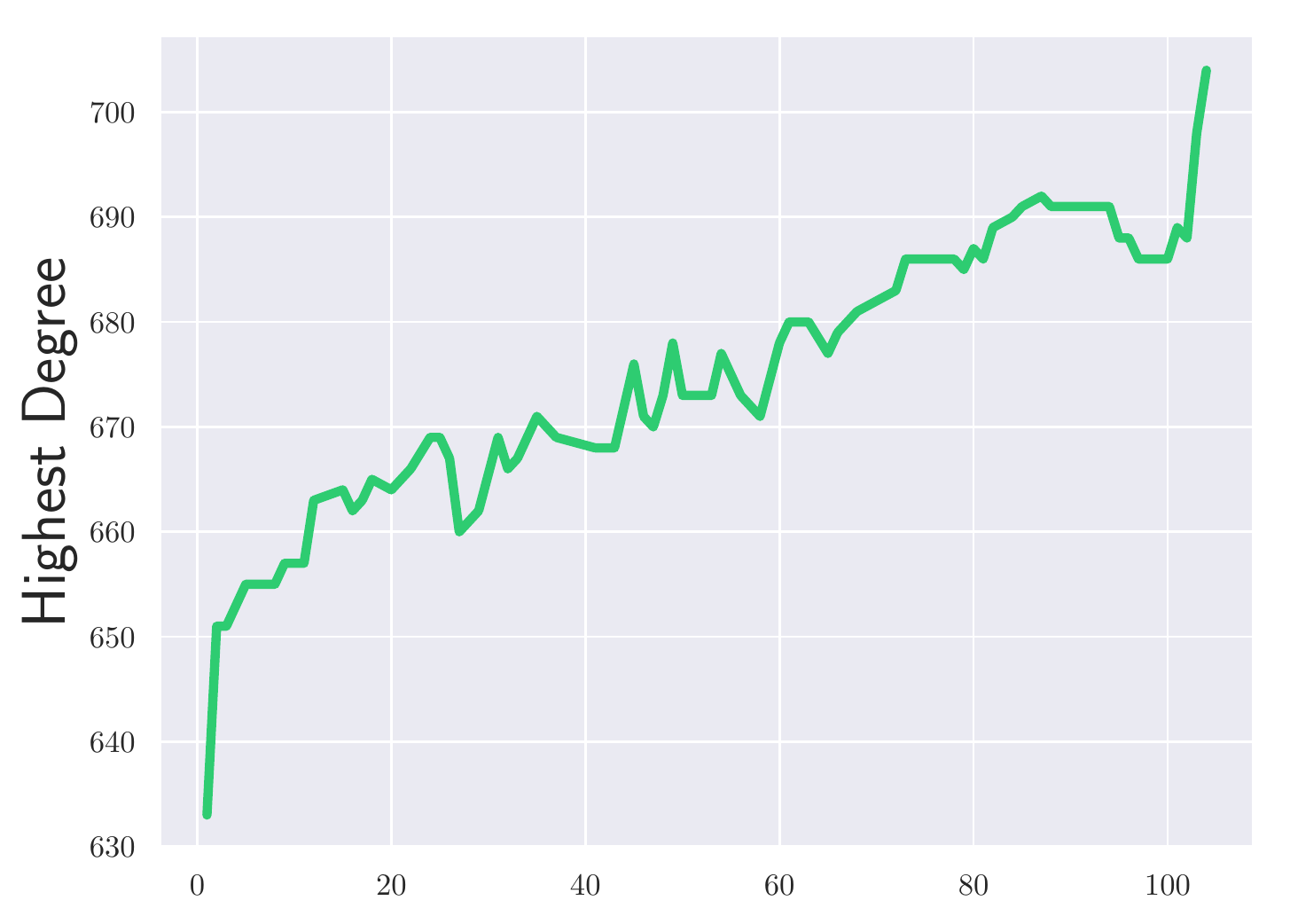}}
     \hfil
    \subfloat{ \includegraphics[width=0.32\textwidth]{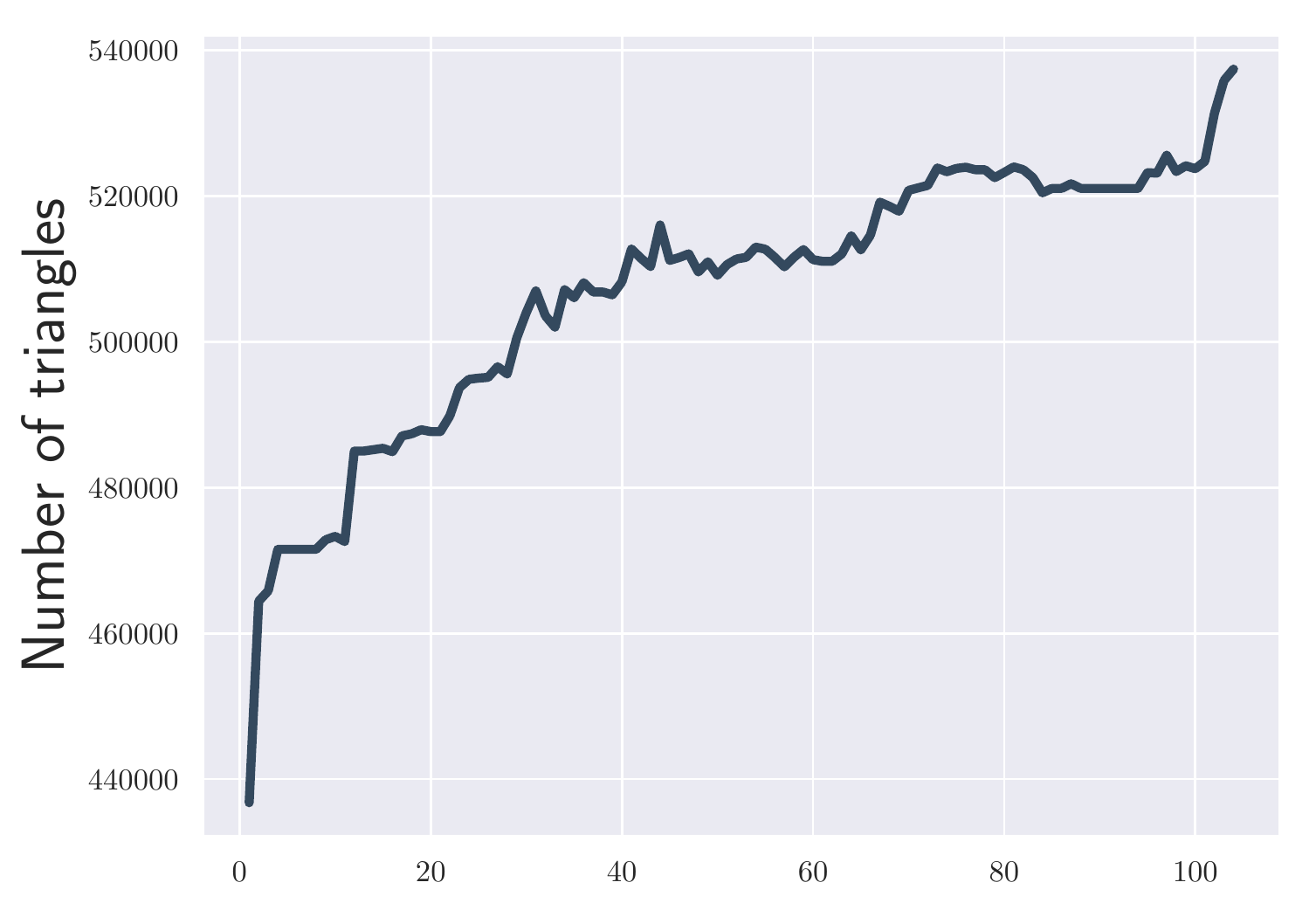}}
     \hfil
    \subfloat{ \includegraphics[width=0.32\textwidth]{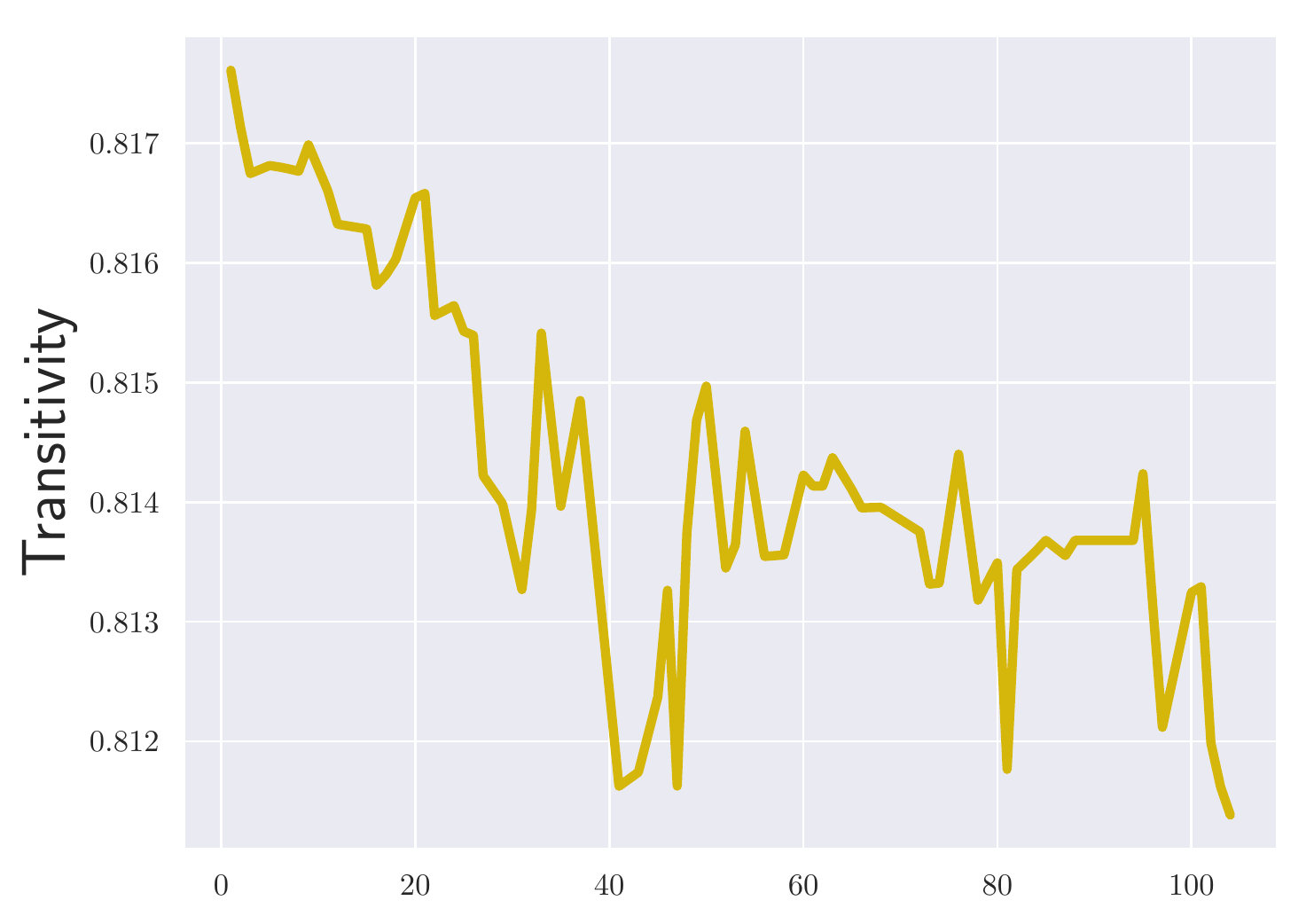}}
    \hfil
    \subfloat{ \includegraphics[width=0.32\textwidth]{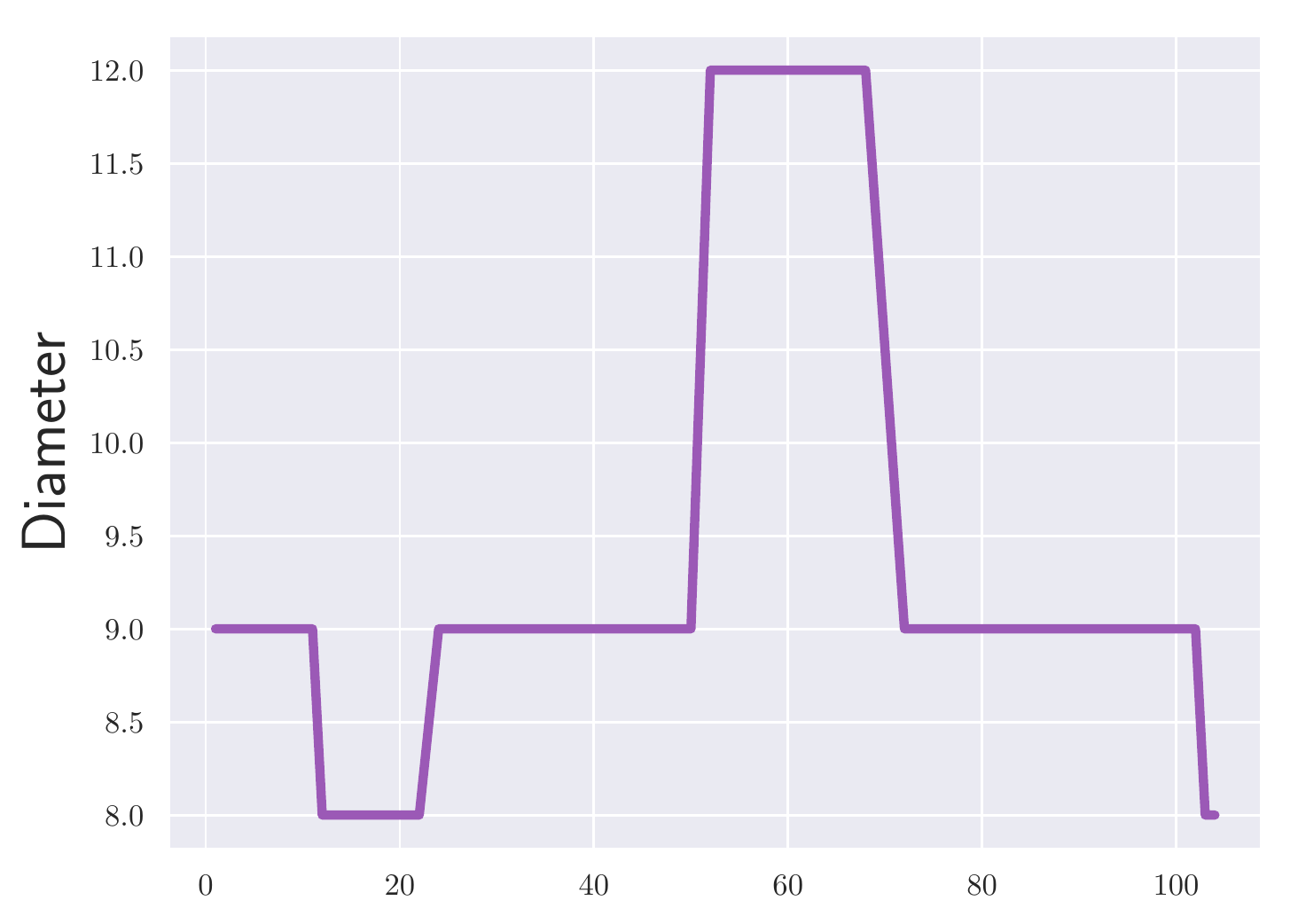}}
     \hfil
    \subfloat{ \includegraphics[width=0.32\textwidth]{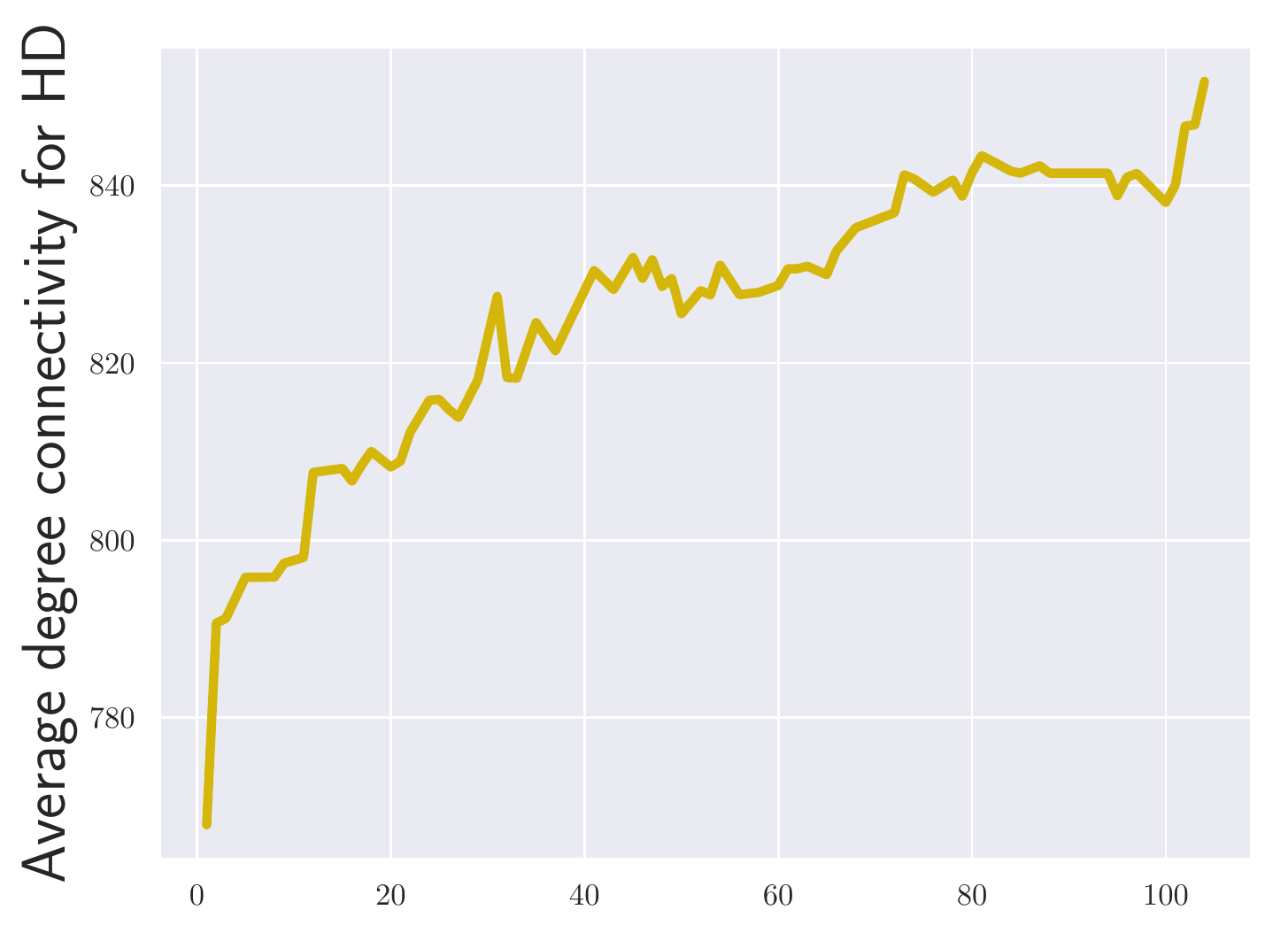}}
     \hfil
    \subfloat{ \includegraphics[width=0.32\textwidth]{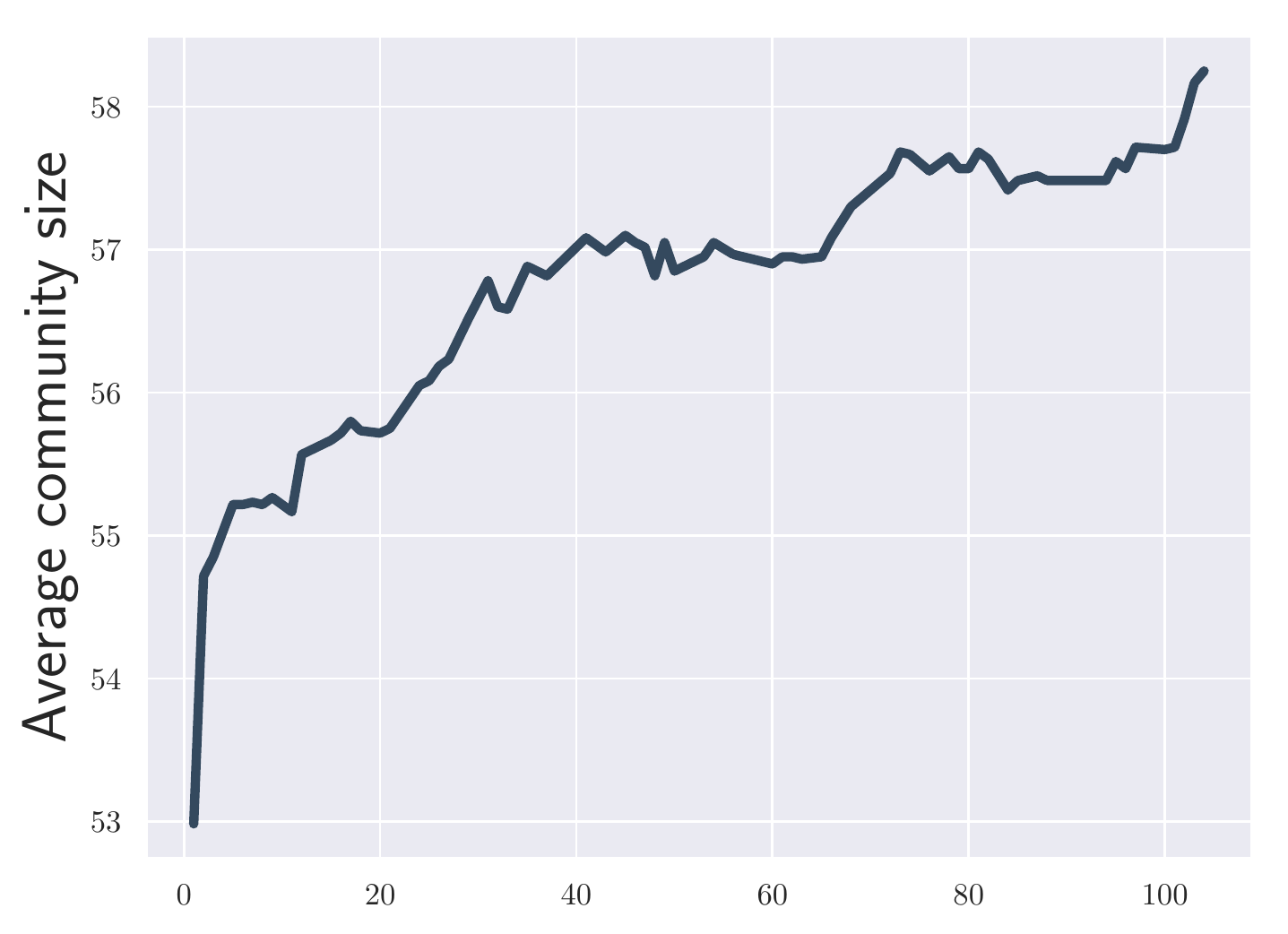}}
  \hfil   
        \caption{Statistics of the temporal graphs over 104 days (Time interval = 6 hours).}
    \label{fig:graph_statistics1}
  \end{adjustbox}
\end{figure*}

\subsection{Gaps}
There are two main reasons for the gaps in the dataset: server errors, and user settings changes. Whenever our data collector detects a server error, it attempts to get that data three times and if it does not succeed, it adds that task to a queue for one more retry before the process termination. user settings changes include disabling the comments of a video and deleting a video. Gaps are marked as NaN and we interpolate the missing data points to eliminate these gaps.

\section{Graph Data}\label{sec:graphdata}

From the temporal data-set, multiple graphs can be induced for analyzing the interaction between the Youtube community. In this paper, we present one of the graphs. The induced graph consists of interaction between user communities through the comments in the Youtube videos. This interaction is captured for multiple time steps. The steps for generation of the temporal graphs are as follows:
 \begin{figure}
   \centering
  \includegraphics[width=1.1\columnwidth]{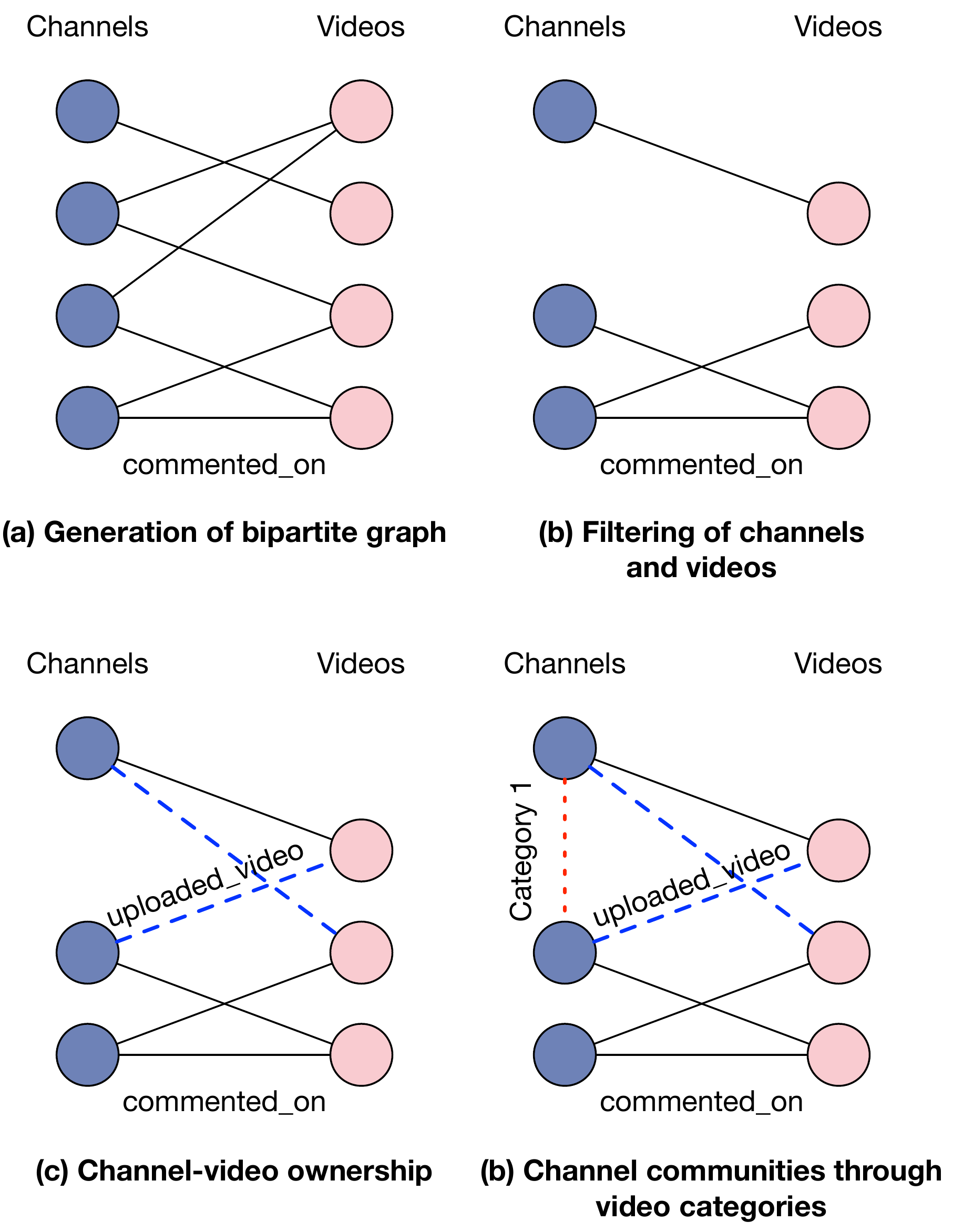}
  \caption{Inducing graph from the dataset.}
  \label{fig:induce}
\end{figure}

\begin{enumerate}
    \item \textbf{Channel-comment-video graph}: First of all, a total of 6,336 channels are monitored every 6 hours to gather their channel, video and comment thread. This is done for a total of 416 time steps. For each time step, we construct a bipartite graph consisting of channels and videos (see Figure \ref{fig:induce} (a)). The edges are added among channels and videos only if the channels have commented in the videos. 
    \item \textbf{Graph filtering}: In the next step, we perform filtering to remove the channels which are not present in the original 6336 channel list (see Figure \ref{fig:induce} (b)). This is done because the channel, video and comment thread data is not collected for the rest of the channels not in the original list of 6336 channels. Moreover, we prune the bipartite graph to remove videos that have less than 1000 view counts, channels that have less than 1000 or more than 10 million subscribers. This is done to remove extreme outlier from the graph.
    \item \textbf{Channel-video ownership addition}: In the next step we add edge between the channels and videos based on the ownership (see Figure \ref{fig:induce} (c)). This is done to keep track of how many videos a channel uploads over a period of time. 
    \item \textbf{Channel-community}: In the final step (see Figure \ref{fig:induce} (d)), we add edges between channels based on category identity. There is a total of 64 categories in the current dataset for the channels. One or more categories are assigned by YouTube to each channel. These categories include film, hobby, lifestyle, music, society, sports, technology, video game, etc.
\end{enumerate}

\subsection{Graph Statistics}
We characterize the induced graph in terms of nodes and edges, average cluster coefficient, number of triangles, graph diameter, and average community size. The statistics are as follows:

\subsubsection*{Nodes and Edges}
The node and edges temporal statistics is presented in figure \ref{fig:graph_statistics1}. The total number of nodes varies from 4,213 to 8,845 from day 1 to day 104. Similarly, the number of edges vary from 456,887 from day 1 to 566,594 to day 104. The number of nodes increases almost linearly over time. The edges are weighted, as multiple channels can have more than one categoryID in common.

\subsubsection*{Average clustering coefficient}
The average clustering coefficient of the weighted graph is calculated as follows \cite{saramaki2007generalizations}:

\begin{subequations}
\begin{align}
C & = \frac{1}{n}\sum_{v \in G} c_v \\
c_u & = \frac{1}{deg(u)(deg(u)-1))} \sum_{uv} (\hat{w}_{uv} \hat{w}_{uw} \hat{w}_{vw})^{1/3}
\end{align}
\end{subequations}
where $\hat{w}_{uv}$ are the edge weights, and these weights are normalized by the maximum weight of the network $\hat{w}_{uv} = w_{uv}/\max(w)$ \cite{hagberg2008exploring}. From figure \ref{fig:graph_statistics1}, it can be noticed that with increasing time steps the average clustering coefficient decreases. More precisely, the average clustering coefficient is 0.3453 at day 1 and 0.1980 at day 104. 

\subsubsection*{Number of Triangles}
For the induced graph the maximum number of triangles present at each time step is presented in Figure \ref{fig:graph_statistics1}. The total number of triangles varies from 436,761 from day 1 to 537,365 from day 104. It can be noticed from the figure that this change in the number of triangles is closely related to the increase in the number of edges over time. 


\subsubsection*{Graph Diameter}
The diameter of the graph is measured by calculating the maximum eccentricity of the graph. the eccentricity of a node $u$ on a graph is defined as the maximum distance from node $u$ to all the other nodes in a graph. The network diameter for the given induced graph is almost constant in most of the time steps. The diameter value in fact only changes among the values 8, 9 and 12. 

\subsubsection*{Transitivity}
The transitivity of the induced graph is defined as follows:
\begin{equation}
  T = 3\frac{\#triangles}{\#triads}  
\end{equation}
It is also known as the fraction of all possible triangles in a graph. From Figure \ref{fig:graph_statistics1}, it can be seen that the transitivity of the induced graph slightly decreases over time. It changes from 0.8176 from day 1 to 0.8114 from day 104. 

\subsubsection*{Highest degree}
The average degree connectivity of a graph is calculated as follows \cite{hagberg2008exploring}:
\begin{equation}
 k_{nn,i}^{w} = \frac{1}{s_i} \sum_{j \in N(i)} w_{ij} k_j
 \end{equation}
 where $s_i$ represents the weighted degree of the node $i$, $W_{ij}$ represents the weight of the edge that links $i$ and $j$, and $N(i)$ represents the neighbors of the node $i$. For the graph the highest  degrees changes from 633 from day 1 to 704 from day 104. Moreover, the actual average connectivity for the respective degree from day 1 is 767.90, and the average connectivity for the degree 704 from day 104 is 851.77.

 \subsubsection*{Average community size}
For the given 6,336 channels, the total number of categories collected is 64. For these 64 communities, the average community size of channels and videos is shown in Figure~\ref{fig:graph_statistics1}.

Overall, we observe that the graph becomes slightly sparser with time as evidenced by the decreasing clustering coefficient and transitivity denoting the lack of strong community formation between the channels and videos over time. However, the average community size increases over time. Thus, many communities spread the influence on other channels and videos to increase community size.

\subsection{Induced Graph Clustering}
\begin{figure}[!htpb]
    \centering
    \begin{adjustbox}{minipage=\linewidth,scale=1}
    \subfloat{\includegraphics[width=0.80\textwidth]{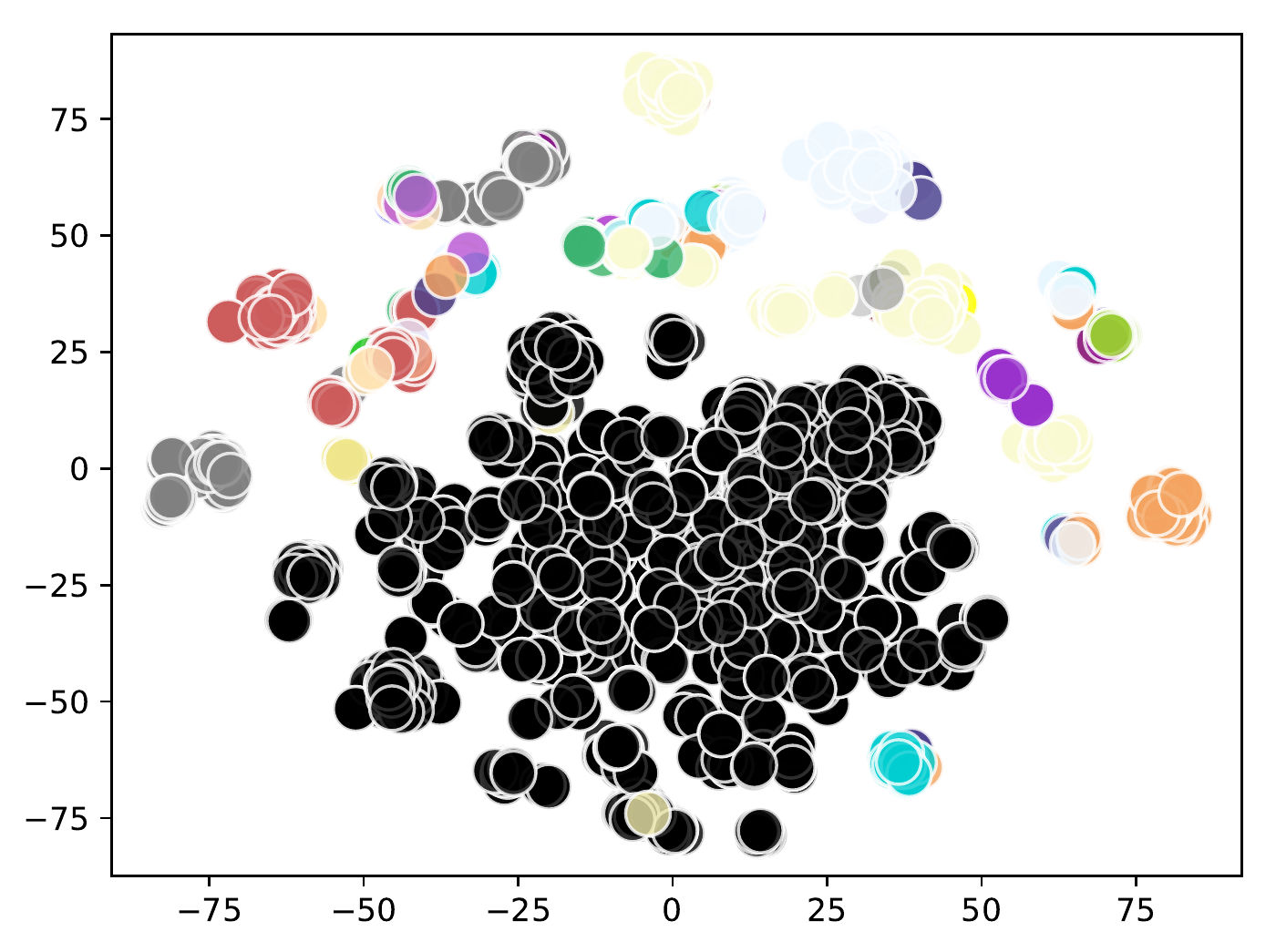}}
    \hfil
    \subfloat{\includegraphics[width=0.80\textwidth]{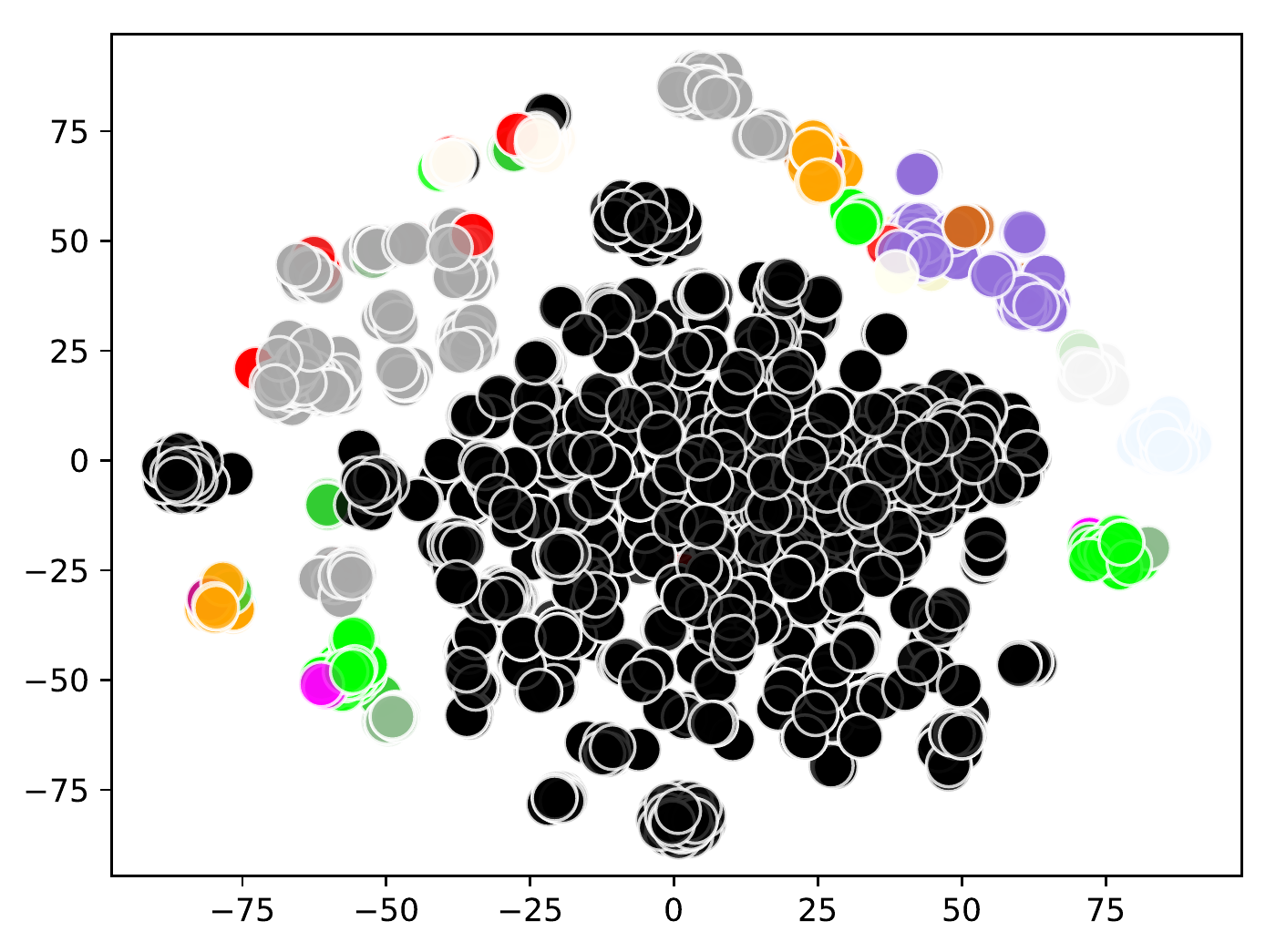}}
    \hfil
    \subfloat{\includegraphics[width=0.80\textwidth]{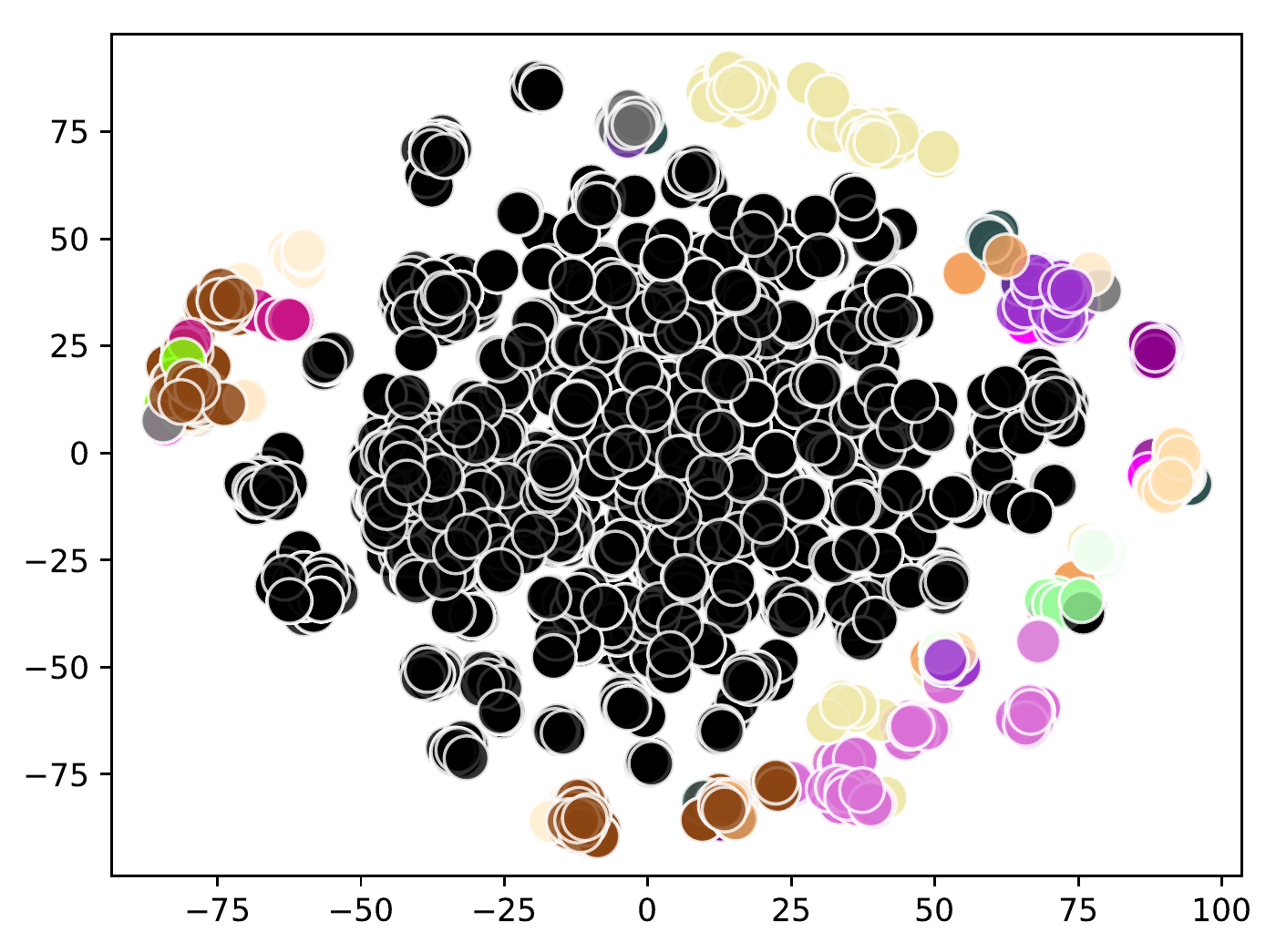}}
       \hfil
    \subfloat{ \includegraphics[width=0.80\textwidth]{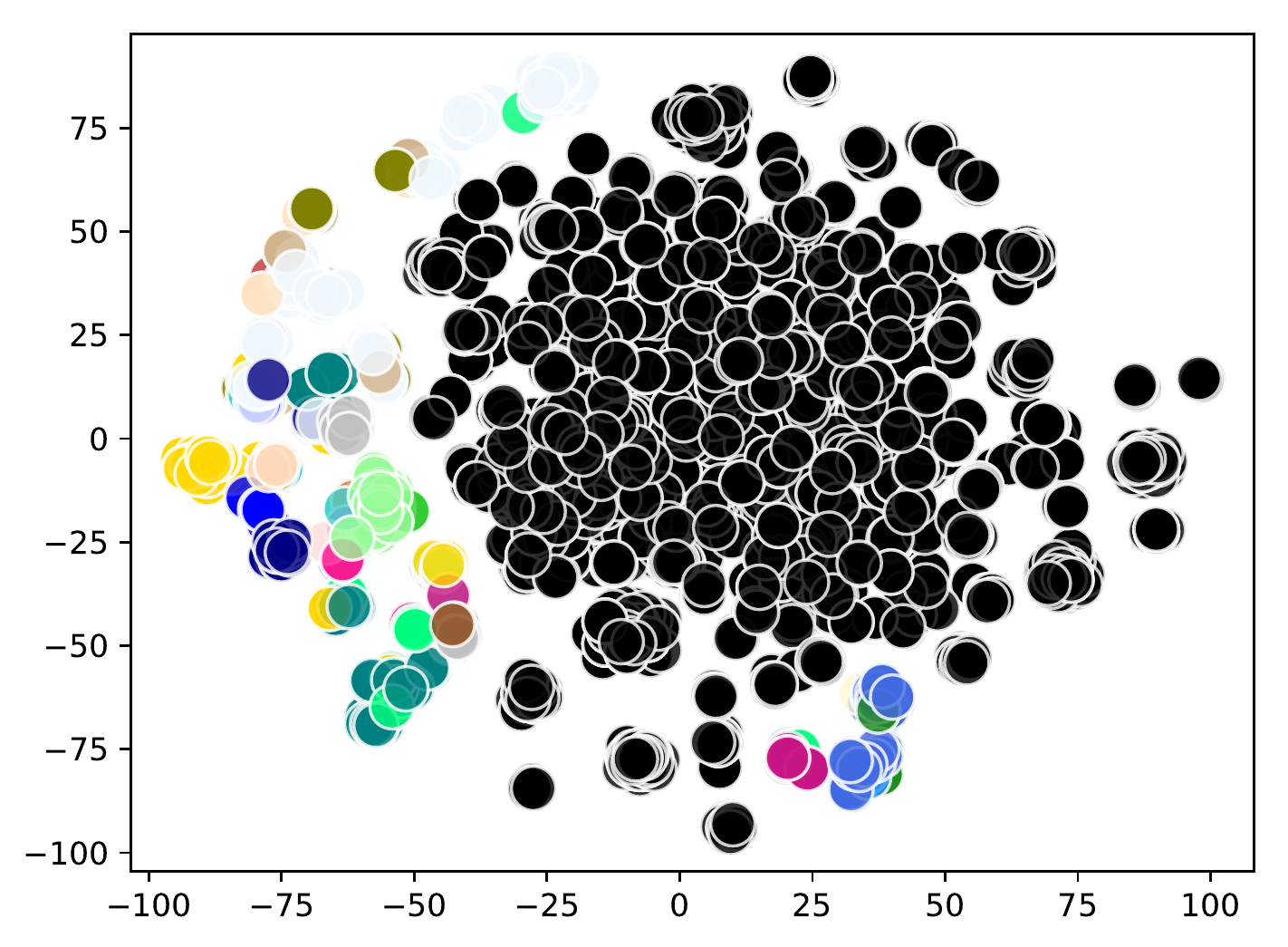}}
     \hfil
    \caption{Community clustering for the temporal graphs (for snapshots 1, 25, 50, and 100.}
    \label{fig:cluster}
\end{adjustbox}
\end{figure}

Figure~\ref{fig:cluster} presents the clustering analysis performed on the induced graphs. We utilized the state-of-the-art static graph embedding algorithms called Structural Deep Network Embedding (SDNE)~\cite{wang2016structural}. SDNE utilizes fully connected deep auto-encoders to embed the given graph to a low-dimensional representation. We used three layers of neural networks in each of the encoder (with the output size of 500, 300 and 128) and decoder (with the output size of 300, 500, and a total number of nodes). The high-dimensional graph is embedded into a vector of size 128. This embedding is then reduced to a size of 2 using t-Distributed Stochastic Neighbour Embedding (t-SNE) \cite{maaten2008visualizing}. They are then plotted with the nodes colored according to the ground truth (categoryIDs). The t-SNE plots for time steps 1, 25, 50, and 100 are shown in Figure~\ref{fig:cluster}. The black colored nodes represent the videos, and the other colored nodes represent channels belonging to different communities. These results show that YoutubeGraph-Dyn captures community migration dynamics on channels.

\subsection{Non-timestamped Data Forecasting}
Predicting the future values of non-timestamped data is an important task for graph learning algorithms in order to properly capture the temporal evolution of the graph. In this section, we compare three different models: autoregressive integrated moving average (ARIMA) models~\cite{contreras2003arima}, long short term memory (LSTM)~\cite{greff2017lstm}, and gated recurrent units (GRU)~\cite{chung2015gated} recurrent neural networks. We evaluate their performance on the prediction of the channel's subscriber count. Unlike other non-timestamped data such as likes, dislikes, and comment count that tends to grow over time, the channel subscriber count presents different trends. Some channels grow their subscribers, other channels lose their subscribers, others keep the same number of subscribers, and others show fluctuations over time.

The ARIMA model is based on statsmodels~\cite{Statsmodels}. We train one model per channel capable of forecasting the next value. The first value is predicted using 70\% of the data. We report the mean square error (MSE) between the actual and predicted values. The $p, d, q$ parameters of the model are individually reported. 

The recurrent neural network (RNN) models are implemented in Keras. Unlike ARIMA, a single RNN model is crated for all channels. We split the dataset in 70\% for training and 30\% for testing. Furthermore, we create $b$ batches of $k$-length to train the models. These models consist of an input LSTM or GRU layer, $l>=0$ hidden LSTM or GRU layers, and a dense output layer with a ReLu activation function. The embedding size of $e$ determines the dimensionality of the output space for each of the layers. We use Adam optimizer with a learning rate of 1e-4, and a mean squared error loss function.

\begin{figure}[!h]
\centering
\vspace{-1em}
  \includegraphics[width=0.48\textwidth]{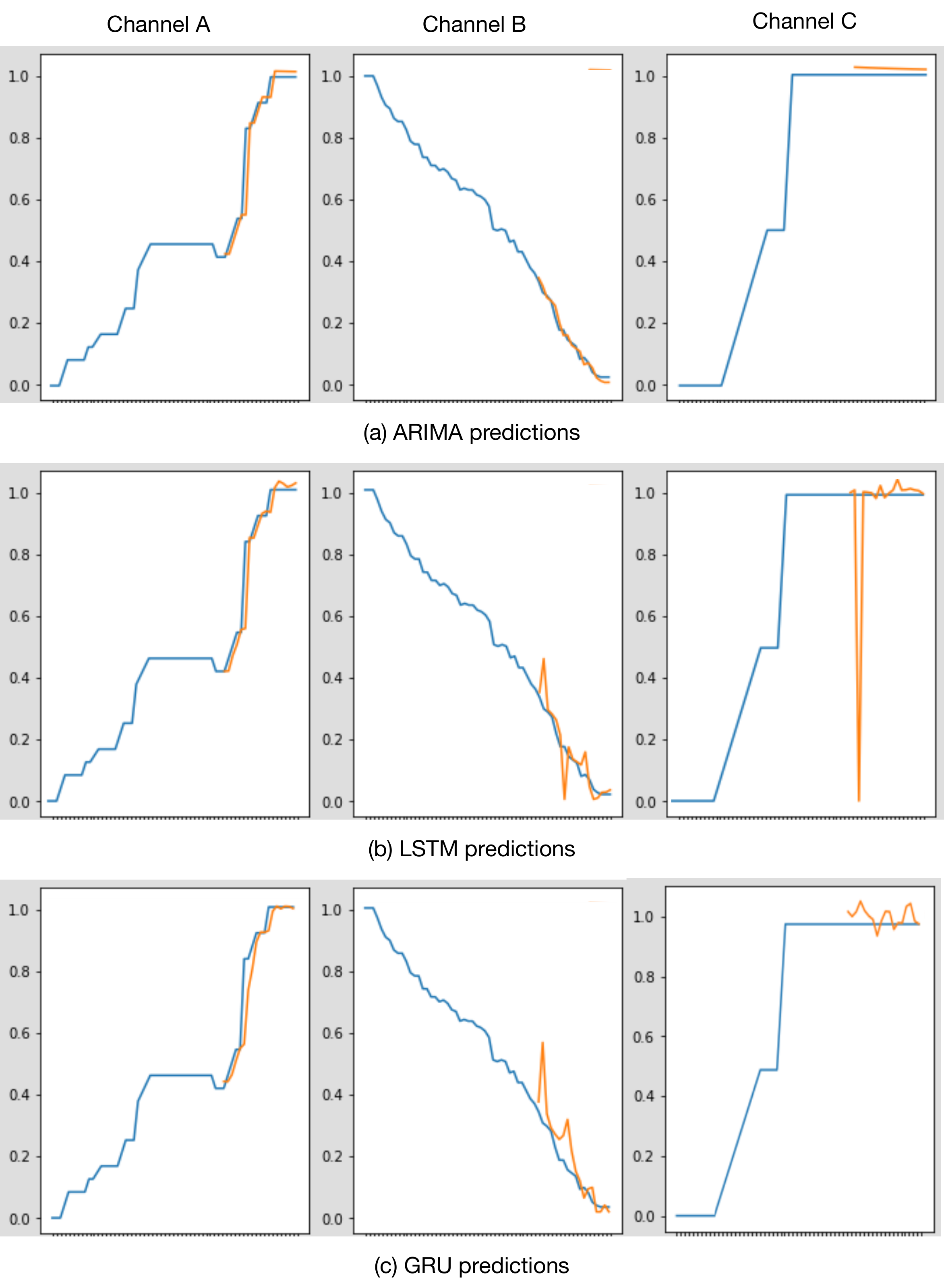}
  \caption{Channel's subscriber count predictions}
  \label{fig:subscriber_arima}
  \vspace{-1em}
\end{figure}

 Figure~\ref{fig:subscriber_arima} shows the normalized subscriber count trends (in blue) for three channels: Channel A, Channel B, and Channel C. Channel A shows a steady increase in subscriber count, followed by a plateau, and finishing with a steady increase of subscribers. Channel B shows a steady decrease in subscribers. And Channel C shows a sharp increase in subscribers and manages to keep those subscribers until the end of the time steps. Figure~\ref{fig:subscriber_arima}(a) shows the ARIMA predictions (in orange) for the three channels.
 
\begin{figure*}
\centering
  \includegraphics[width=1\textwidth]{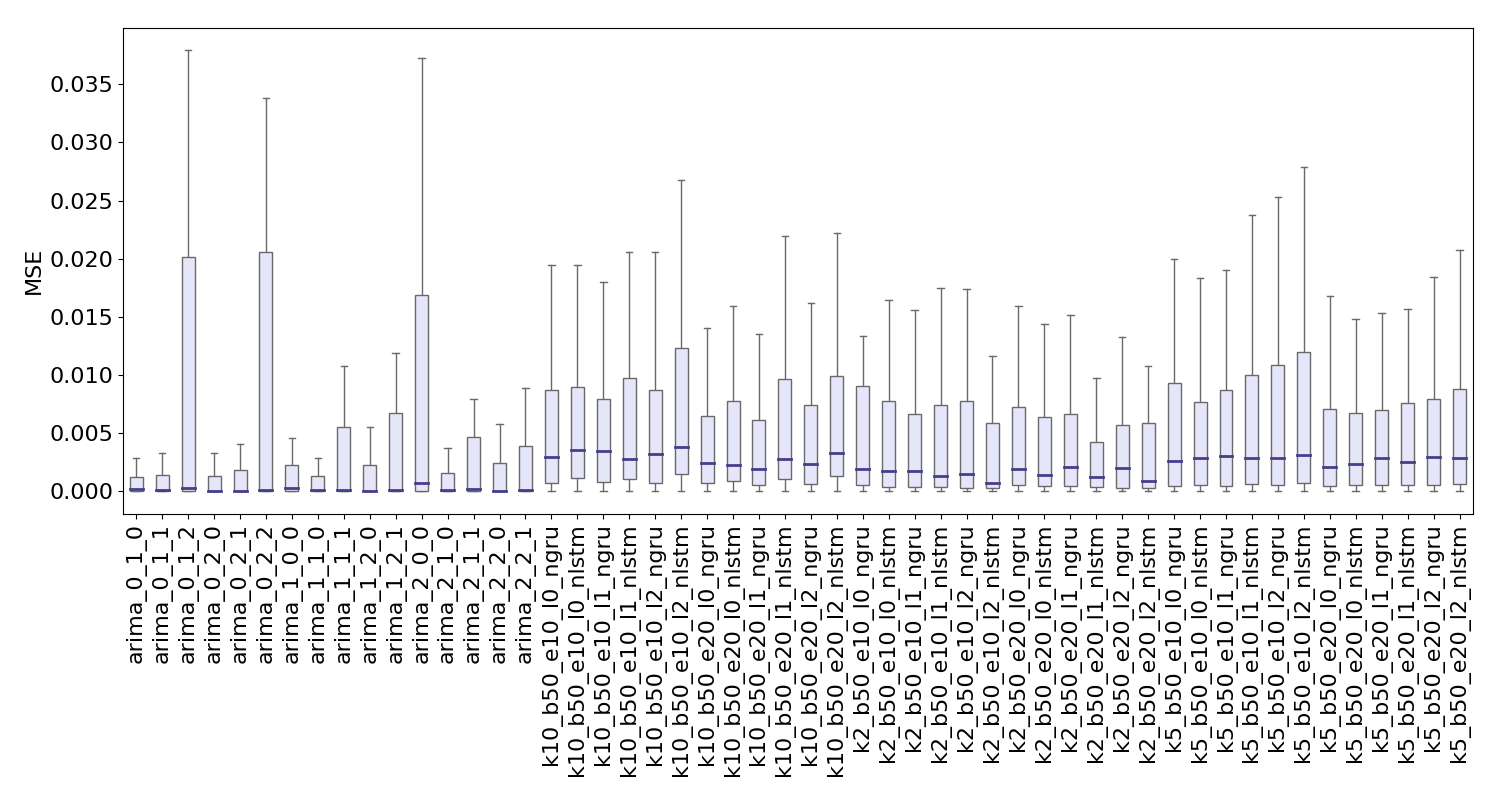}
  \caption{Exploration of hyperparameters for ARIMA, LSTM, and GRU models.}
  \label{fig:all_count_results}
\end{figure*}

 Figure~\ref{fig:subscriber_arima}(b) shows the LSTM predictions, and Figure~\ref{fig:subscriber_arima}(c) shows the GRU predictions using a lookback of 5, batch size of 50, and embedding size of 10. The three models succeed at predicting the general trends of the three channels. In particular, the ARIMA models provide a more accurate prediction of the next subscriber count values for the three channels. The LSTM and GRU models give good predictions for Channel A but are less accurate for channels B and C. Figure~\ref{fig:all_count_results} shows the distribution of MSE for various subscriber count predictor model parameterizations. The tick represents the median MSE for all 6,342 channels.

The \texttt{arima\_x\_y\_z} models are the configurations for different $x=p$, $y=d$, $z=q$ values. For example, the \texttt{arima\_0\_1\_0} represents an ARIMA model parameterized as $p=0$, $d=1$, $q=0$. On the other hand, the RNN models indicate the $k$-length of $b$ batches with an $e$ embedding size and $l$ hidden layers of type $gru$ or $lstm$. For example, \texttt{k10\_b50\_e10\_l0\_ngru} represents a GRU model with no hidden layers, an embedding size of 10, a batch size of 50 with 10-length samples per batch.

In general, the ARIMA models achieve a smaller MSE compared to the RNN models. However, the RNN models generalize the data better as a single RNN model is trained for all 6,342 channels; an individual ARIMA model is trained per channel. For the RNN models, the $k=2$ configurations provide the best predictions. Both the ARIMA and RNN models can be used to forecast YoutubeGraph-Dyn's non-timestamped data.

\section{Discussion}\label{sec:discussion}
In this section, we summarize the limitations of YoutubeGraph-Dyn and provide our perspective on possible extensions and future research. First, the channel-comment-video temporal graph was the main focus of this paper. This is motivated by the need of having better datasets for dynamic graph learning. However, YoutubeGraph-Dyn can be used to induce other types of graphs. Second, YoutubeGraph-Dyn provides finer time granularity compared to similar datasets; we showed that 6-hour intervals are sufficient to capture channel community migration dynamics. However, YoutubeGraph-Dyn's time granularity may not be adequate to capture faster dynamics in the network. Third, YoutubeGraph-Dyn is limited to a single social network (YouTube). We anticipate that future dynamic graph learning research will require richer temporal graphs that go beyond a single network. 

\subsection{Other types of induced graphs}
In this paper, we focused on channel-comment-video graphs. However, there are other types of graphs that can be induced from the raw data. Some of the graphs that can be induced are as follows:
\begin{itemize}[leftmargin=0.10in, labelsep=0.05in, itemsep=0.02in, parsep=0in]
    \item \textbf{User-comment-user graph}: A user-comment-user graph can be built from comment threads when two users interact through a reply. This graph could be used, for example, to study the structure and dynamics of influencers.
    \item  \textbf{Channel-category-channel graph}: A channel-category-channel graph can be built when a user uploads a video which is common with other channels. Since the channels can have multiple common categories, weighted edges can be added among the channels. As channels will have videos uploaded under different categories, this graph will help to study the trend followed by the channels. Moreover, it will also allow us to study the evolution of the category community.
    \item  \textbf{User-category-user graph}: A user-category-user graph can be constructed by adding edges between users who comment on videos with similar video category. Unlike user-comment-user graph, where the edge between users are added when they interact with a reply, a user-category-user graph will not require interaction. Moreover, unlike the channel-category-channel graph, the users may not have any video uploaded. This graph will help us in analyzing the evolution of users community with respect to the video category.
    \item  \textbf{Subscriber count ego network}: To study the popularity of the channels, a subscriber count ego network may be created. Since subscriber count is continuous value, it has to be binned first. A logarithmic scale of base 10 may appropriately capture the range of subscribers. After binning the subscriber count, the channels with a specific number of subscribers are added to the ego network. Moreover, an edge between the channels is added based on the category of videos uploaded by them. This graph will allow us to study how various unpopular channels evolved to become popular by publishing videos in topics that are most popular at a particular period of time. 
    \item  \textbf{Channel:video-sentiment-user graph}: The comments provided by the users in the videos may be processed to perform sentiment analysis and ranked based on positive or negative sentiments with weights ranging from 1 to 10, 1 being most negative and 10 being most positive. Based on this a channel:video-sentiment-user graph may be created, where there are two types of nodes a channel:video node and a user node. The weighted edge between them is based on the sentiment score. This graph will allow us to study the overall standing of channels in the user community.  
\end{itemize}

\subsection{Intra-hour time granularity}
The YouTube API restricts the number of requests (i.e., credits) that one account can perform per hour. Therefore, the work can be distributed among various accounts to either increase the volume of channels, comments, and videos or to take intra-hour snapshots of the data. In our case, three accounts (one per author) were sufficient to take snapshots of the 6,342 channels including all their videos and comments every 6 hours. The motivation behind intra-hour snapshots is to capture faster network dynamics such as virality. We anticipate intra-hour granularity to provide interesting dynamics for new algorithms.

\subsection{Multi-network temporal graphs}
In many cases, users have a presence in several social networks such as YouTube, Twitter, and Instagram. Analyzing multi-network interactions is an interesting area of research that requires multi-network temporal graphs. However, the methodology to create such a dataset must be first developed and we leave that for future work.

\section{Conclusion}\label{sec:conclusion}
YoutubeGraph-Dyn takes a first step in filling the gap that exists in datasets that capture the evolution of graphs. Our approach, based on non-timestamped data from the YouTube API, allowed us to create a channel-comment-video time evolving graph. Using SDNE, a state-of-the-art graph embedding algorithm, we demonstrated that YoutubeGraph-Dyn captures community migration dynamics in YouTube channels. YoutubeGraph-Dyn provides 106 intra-day graph snapshots taken every 6 hours resulting in a total of 416 time steps. A key differentiator of YoutubeGraph-Dyn is the encoding of various types of relationships and multiple attributes in the form of word embeddings and integers. We showed that RNNs and ARIMA models can be used to predict non-timestamped data such as subscriber count.


\bibliographystyle{ACM-Reference-Format}

\bibliography{bibliography}

\end{document}